\documentclass[modern]{aastex61}
\usepackage{natbib}

\received{}
\revised{}
\accepted{\today}
\submitjournal{ApJ}

\shorttitle{Microwave Observations of a Solar Limb Flare }
\shortauthors{Gary et al.}

\begin{document}

\title{Microwave and Hard X-Ray Observations of the 2017 September 10 Solar Limb Flare}

\correspondingauthor{Dale E. Gary}
\email{dgary@njit.edu}

\author[0000-0003-2520-8396]{Dale E. Gary}
\affil{Center for Solar-Terrestrial Research, New Jersey Institute of Technology,
323 M L King Jr Blvd, Newark, NJ 07102-1982, USA}

\author[0000-0002-0660-3350]{Bin Chen}
\affil{Center for Solar-Terrestrial Research, New Jersey Institute of Technology,
323 M L King Jr Blvd, Newark, NJ 07102-1982, USA}

\author{Brian R. Dennis}
\affil{Solar Physics Laboratory, NASA Goddard Space Flight Center, Greenbelt, MD 20771, USA}

\author[0000-0001-5557-2100]{Gregory D. Fleishman}
\affil{Center for Solar-Terrestrial Research, New Jersey Institute of Technology,
323 M L King Jr Blvd, Newark, NJ 07102-1982, USA}

\author{Gordon J. Hurford}
\affil{Space Sciences Laboratory, University of California, Berkeley, CA 94720, USA} 

\author{S\"am Krucker}
\affil{Space Sciences Laboratory, University of California, Berkeley, CA 94720, USA} 
\affil{Institute for Data Science, University of Applied Sciences and Arts Northwestern Switzerland, 5210 Windisch, Switzerland}

\author{James M. McTiernan}
\affil{Space Sciences Laboratory, University of California, Berkeley, CA 94720, USA} 

\author[0000-0003-2846-2453]{Gelu M. Nita}
\affil{Center for Solar-Terrestrial Research, New Jersey Institute of Technology,
323 M L King Jr Blvd, Newark, NJ 07102-1982, USA}

\author{Albert Y. Shih}
\affil{Solar Physics Laboratory, NASA Goddard Space Flight Center, Greenbelt, MD 20771, USA}

\author{Stephen M. White}
\affil{Space Vehicles Directorate, Air Force Research Laboratory} 

\author[0000-0003-2872-2614]{Sijie Yu}
\affil{Center for Solar-Terrestrial Research, New Jersey Institute of Technology,
323 M L King Jr Blvd, Newark, NJ 07102-1982, USA}



\begin{abstract}

We report the first science results from the newly completed Expanded Owens Valley Solar Array (EOVSA), which obtained excellent microwave imaging spectroscopy observations of SOL2017-09-10, a classic partially-occulted solar limb flare associated with an erupting flux rope. This event is also well-covered by the Reuven Ramaty High Energy Solar Spectroscopic Imager (RHESSI) in hard X-rays (HXRs). We present an overview of this event focusing on microwave and HXR data, both associated with high-energy nonthermal electrons, and discuss them within the context of the flare geometry and evolution revealed by extreme ultraviolet (EUV) observations from the Atmospheric Imaging Assembly aboard the Solar Dynamics Observatory (SDO/AIA).  The EOVSA and RHESSI data reveal the evolving spatial and energy distribution of high-energy electrons throughout the entire flaring region. The results suggest that the microwave and HXR sources largely arise from a common nonthermal electron population, although the microwave imaging spectroscopy provides information over a much larger volume of the corona. 
%
%
%
%
%
%

\end{abstract}




\section{Introduction} \label{sec:intro}

It has long been recognized that microwave (MW) and hard X-ray (HXR) observations of solar flares are highly complementary, both emissions arising from high-energy electrons accelerated during the energy release process.  Although these two emissions often have extremely similar lightcurves during the impulsive phase \citep{1988SoPh..118...49D}, important differences remain, namely that the MW-producing gyrosynchrotron emission arises mainly from a trapped population of electrons spiraling in coronal magnetic loops, while the HXR emission is dominated by bremsstrahlung from precipitating electrons escaping to the footpoints of those same loops.  HXR images and spectra often tell a more complex story, however, revealing both a super-hot thermal component in the corona and sometimes, especially in cases of occulted or partially-occulted limb flares, a nonthermal ``above-the-looptop" coronal source \citep[e.g.,][]{1994Natur.371..495M,2008ApJ...673.1181K}.

For more than two decades, MW studies of solar flares have been dominated by data from the solar-dedicated Nobeyama Radioheliograph \citep[NoRH;][]{1994IEEEP..82..705N} taken at two fixed frequencies, 17 and 34~GHz. NoRH was designed to operate at optically-thin frequencies well above the typically 5-10~GHz peak of the MW spectrum \citep{1975SoPh...44..155G,2004ApJ...605..528N}, where the interpretation of the emission is expected to be relatively simple.  The MW emission at these high frequencies comes mainly from regions of high magnetic field strengths, and hence are likely to be relatively small, compact loops, or the footpoints of larger loops \citep[e.g.,][]{1997ApJ...489..976N,1997SoPh..173..319H}.  However, large loop-top sources due to efficient trapping of the nonthermal electrons have been reported \citep[e.g.,][]{2002ApJ...580L.185M}. Observations by the Owens Valley Solar Array (OVSA, \citealt{1994ApJ...433..379W}), and occasionally by the Very Large Array (VLA, e.g., \citealt{1990ApJ...358..654S}), revealed a richer range of phenomena that could be exploited with microwave imaging spectroscopy---the use of data with simultaneous high spatial, spectral, and temporal resolution over a broad frequency range. In particular, they revealed evidence for extremely large MW sources at lower frequencies \citep{1994SoPh..152..409L,1994ApJ...433..875K,Fl_etal_2017,2018ApJ...852...32K}, where emission from energetic electrons in regions of weak magnetic field becomes visible, as well as purely thermal microwave sources \citep{Gary_Hurford_1989,2015ApJ...802..122F, 2017NatAs...1E..85W}. 
%
%

Recognizing the potential for microwave imaging spectroscopy, a concept for a new solar-dedicated array (the Frequency-Agile Solar Radiotelescope, or FASR, \citealt{2004ASSL..314.....G,2005ASPC..345..142B}) was developed to provide the capabilities needed to exploit this technique.  Although FASR has not yet been realized, its design concepts have been applied in the creation of a smaller, demonstrator array called the Expanded Owens Valley Solar Array (EOVSA) that has been fully operational since April 2017. This paper describes the first example of microwave imaging spectroscopy from EOVSA, and demonstrates that it has achieved the performance expected from earlier simulations \citep{2013SoPh..288..549G}. We choose for this first report an extremely well observed, partially-occulted limb flare associated with an erupting flux rope, seen in profile in extreme ultra-violet (EUV) and HXR emissions. This unique combination of data fully captures the event within the framework of the standard solar flare model, also known as the CSHKP model \citep{1964NASSP..50..451C,1966Natur.211..695S,1974SoPh...34..323H,1976SoPh...50...85K}, but in addition, with the unprecedented MW spectral imaging, reveals new information about the extent of highly energetic (100s of keV to MeV) electrons within that framework that has heretofore been hidden.
%
%

\section{Observations} \label{sec:observations}

We present EOVSA observations of the X8.2 flare, SOL2017-09-10, that was partially-occulted by the west limb and peaked at around 16:00 UT. It continued to produce emissions in MW, extreme ultraviolet (EUV), X-rays, and $\gamma$-rays for many hours. The event was well observed in MW by EOVSA, in EUV by the Atmospheric Imaging Assembly (AIA, \citealt{2012SoPh..275...17L}) on the Solar Dynamics Observatory (SDO), in HXR by the Reuven Ramaty High-Energy Solar Spectroscopic Imager (RHESSI, \citealt{2002SoPh..210....3L}) and the Fermi Gamma-ray Burst Monitor (GBM, \citealt{2009ApJ...702..791M}), and in $\gamma$-rays by the Fermi Large Area Telescope (LAT, \citealt{2009ApJ...697.1071A}).  The event has received considerable attention in the literature in the few months since it occurred (e.g. \citealt{2018ApJ...854..122W,2018ApJ...853L..18Y,2018ApJ...853L..15L,2018ApJ...855...74L,2018arXiv180307654O,2018ApJ...853..178D}).
%
%

\begin{figure}[ht]    
   \centerline{\includegraphics[width=1.0\columnwidth,clip=]{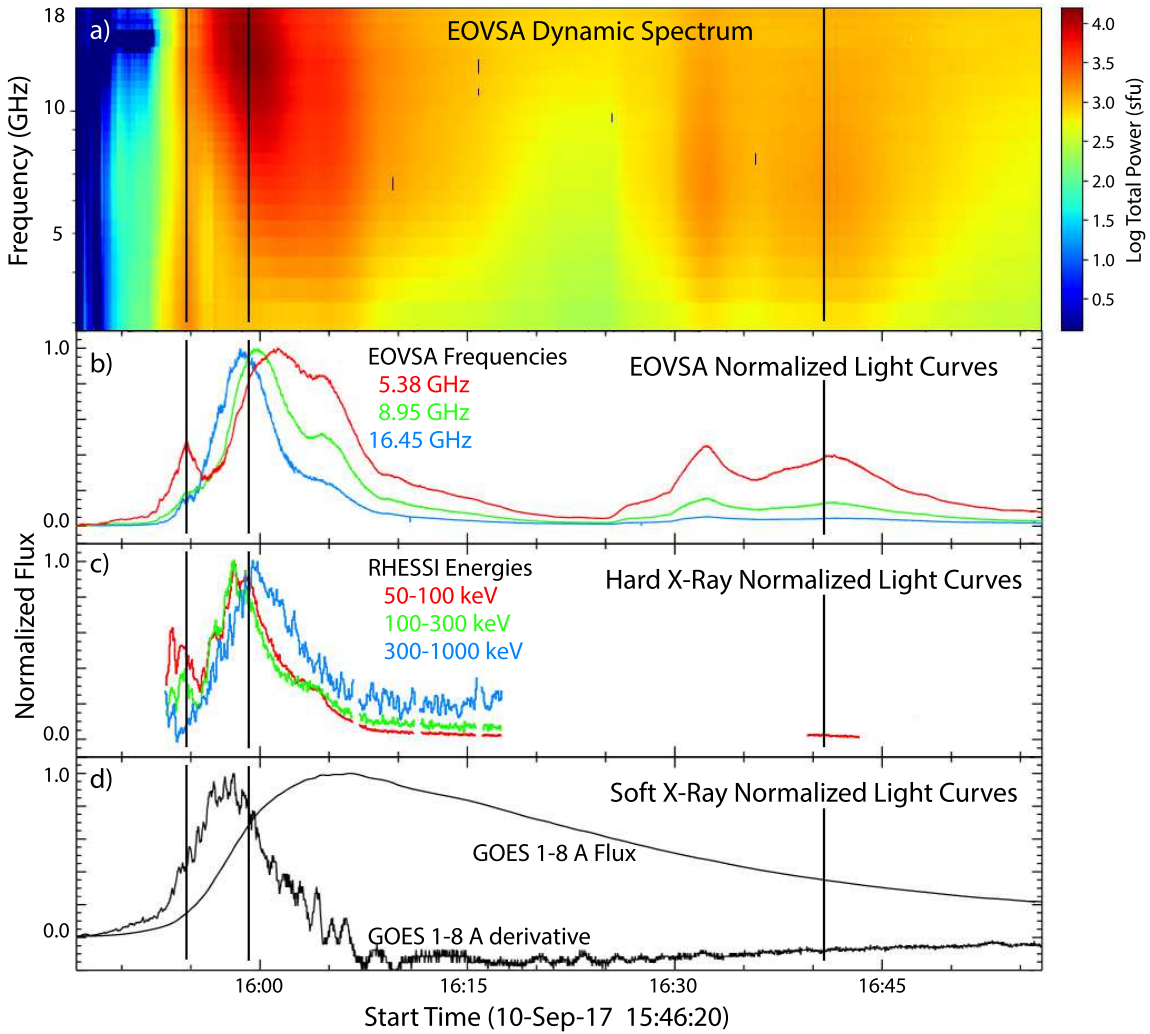}
              } 
              \caption{Dynamic MW spectrum and normalized lightcurves of the first $\sim$1~hour of the event at different wavelengths. (a) The EOVSA total power dynamic spectrum from 2.5-18 GHz, with colors representing the flux density in sfu, as shown in the color bar at right. (b) Normalized time profiles of the MW emission at three frequencies. (c) Normalized time profiles of RHESSI HXR counts, with a gap due to passage of the spacecraft through the South Atlantic Anomaly. (d) Normalized GOES 1-8 \AA\ flux and time derivative.  The vertical lines mark three times discussed in more detail in the text.
                      }
   \label{fig1}
\end{figure}

Figure~\ref{fig1} shows an overview of spatially-integrated lightcurves for emission at different wavelengths, and the total power microwave dynamic spectrum from EOVSA.  The lightcurves are normalized to unity to emphasize the differing peak times. The vertical black lines mark three specific times in the event that we focus on: (1) an early impulsive peak near 15:54 UT ($t_1$) that has a nearly flat MW spectrum, (2) the peak time near 16:00 UT ($t_2$) that has a steeply rising MW spectrum, and (3) a time near 16:41 UT ($t_3$), when RHESSI resumed solar observations after passage through the South Atlantic Anomaly.  Of particular interest is the comparison of peak times.  The GOES 1-8 \AA\ flux derivative in Figure~\ref{fig1}d peaks around 15:57 UT, close to the time of the RHESSI 50-300 keV lightcurves at 15:58 UT (Figure~\ref{fig1}c). However, the peak in the higher energy 300-1000 keV RHESSI lightcurve is delayed to 16:00 UT. 
This implies a progressive increase in energy of the accelerated particles during this phase of the event.  For a static source, such an evolution of particle energy would be expected to shift the MW peak frequency to higher frequencies, leading to higher frequencies peaking later \citep[e.g.][]{1985ARA&A..23..169D}.  However, as shown in Figure~\ref{fig1}b, the delay in peak time is opposite to this expectation, with higher frequencies peaking earlier (15:58:50~UT at 18~GHz) and lower frequencies peaking later (16:01:30~UT at 5.4~GHz). As the MW images will show, this progressive delay with frequency is due to a relatively slow evolution of the entire MW-emitting source region from low coronal heights with higher magnetic field strength toward greater heights and lower magnetic field strength. This spatial evolution thus leads to a more complicated total power (spatially integrated) spectral evolution, in this case actually inverting the expected delay with frequency.

\begin{figure}[ht]    
   \centerline{\includegraphics[width=0.85\columnwidth,clip=]{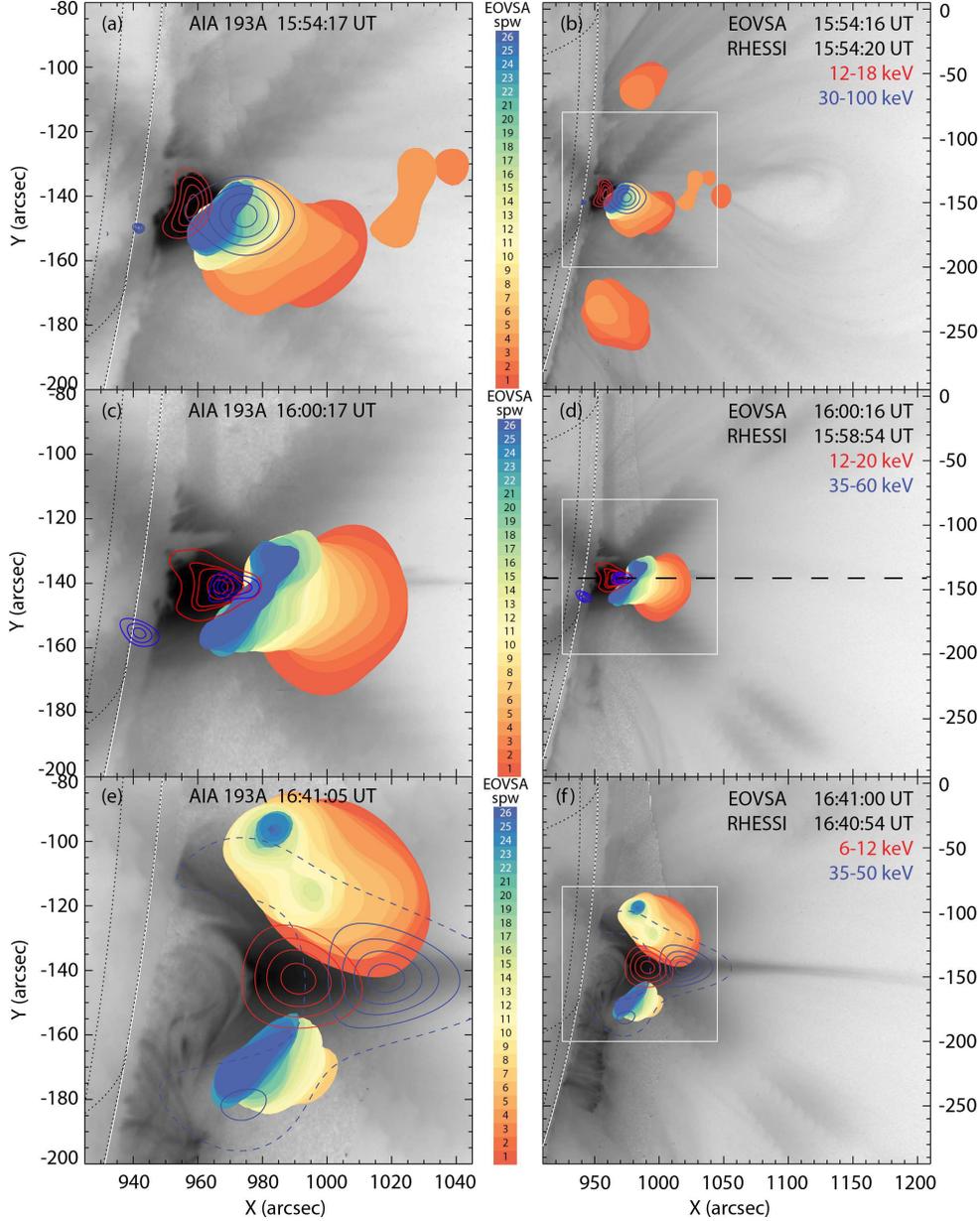}
              }
              \caption{Comparison of AIA, RHESSI, and EOVSA images at the three times marked in Fig.~\ref{fig1}. Each image shows the corresponding AIA 193~\AA\ image (in reverse grayscale of log intensity) superposed with filled 50\% contours of EOVSA MW emission at 26 spectral windows, with hues shown in the color bar. RHESSI HXR 30, 50, 70, and 90\% contours are also superposed for two energy ranges.  (a) Zoomed in ($2\times2$ arcmin) field of view (FOV) of the limb flare near 15:54 UT. (b) Same as (a), but showing a larger $5\times5$ arcmin FOV.  This view shows additional low-frequency MW sources flanking the main source to the north and south.  The white box outlines the area shown in (a). (c) Same as (a), for the peak time near 16:00 UT, except RHESSI 12-20~keV contours are 40, 55, 60, 75 and 90\%. (d) $5\times5$ arcmin FOV corresponding to (c). The horizontal dashed line marks the position of the cut used for the height-time plots of Figs~\ref{fig3} and \ref{fig4}. (e) Same as (a), for a time near 16:41 UT in the decay phase. (f) $5\times5$ arcmin FOV corresponding to (e). The dashed contour in (e) and (f) is the 10\% contour for the RHESSI 35--50~keV image.}
\label{fig2}
\end{figure}

As described in \citet{2018SoPh...xx.xxxG} (see also \citealt{2016JAI.....541009N}), EOVSA is designed to observe at hundreds of frequency channels spread over 34 spectral windows of 500~MHz bandwidth over the 1--18~GHz frequency range, covering the entire spectrum in 1 s.  At the time of these observations, for reasons discussed in that paper, a high-pass filter was in place on each antenna to limit the observations to 2.5--18~GHz in 134 frequencies spread over 31 spectral windows, with the width of each window limited to 160~MHz.  The lowest spectral window is as yet uncalibrated, leaving 30 usable spectral windows.  For simplicity in this first-results paper, we combine the frequency channels in each spectral window, to provide imaging at 30 equally-spaced frequencies ranging from 3.4--18~GHz, with center frequencies $f_{\rm GHz} = 2.92 + n/2$, where $n$ is the spectral window (spw) number from 1-30.  The EOVSA images in spectral windows 1-26 for the three times marked in Figure~\ref{fig1} are shown in Figure~\ref{fig2} as filled 50\% contours overlaid on AIA 193~\AA\ images. The nominal full-width-half-max (FWHM) spatial resolution of these observations is elliptical, with major axis $113.7\arcsec/f_{\rm GHz}$ and minor axis $53.0\arcsec/f_{\rm GHz}$.  During the CLEAN process a circular restoring beam was used of FWHM $89.7\arcsec/f_{\rm GHz}$ for frequencies up to 14.9~GHz, while the size was fixed at 6\arcsec above 15~GHz.  Thus, the frequency range of 3.4--18~GHz corresponds to a restored range of 25.7\arcsec--6\arcsec.

\subsection{EOVSA Source Morphology}
At time $t_1$ shown in Fig.~\ref{fig2}a,b, the EOVSA source at lower frequencies is complex, consisting of a bright central source located well above the bright AIA loops, flanked by two more-distant sources associated with the legs of a much larger loop that appears to be associated with the coronal mass ejection (CME).  In addition, sources at the lowest few frequencies in Fig.~\ref{fig2}b appear distributed along a line connecting the AIA bright loops with a rapidly expanding, tear-drop-shaped cavity seen faintly in 193~\AA. 
In the standard solar flare model, the cavity would be identified with a rising flux rope \citep{2018ApJ...855...74L} and the line connecting it to the lower, bright loops might be identified as a signature of the reconnecting current sheet \citep{2018ApJ...854..122W}, although to avoid over-interpretation we will refer to it as a ``plasma sheet".  The emission at different frequencies in the bright central source, seen most clearly in Fig.~\ref{fig2}a, shows a clear dispersion in height, with the highest-frequency source being lowest and most compact, but still lying well above the bright AIA loops.  This also agrees well with the standard solar flare model, in which the MW emission comes from the most recently closed loops that contain newly accelerated electrons.  However, the emission appears to be more confined to the loop tops than would be expected in the simplest interpretation of the standard model \citep[e.g.][]{1997ApJ...480..825A}. One way to account for the confinement is to invoke a high mirror ratio in the initially collapsing loops \citep[e.g.][]{1998ApJ...505..418F,2004A&A...419.1159K}, but it is likely that turbulence and wave-particle interactions also play a role in mediating the trapping. The dispersion in height with frequency mainly reflects the fall-off of coronal magnetic field strength with height, as discussed in section~\ref{sec:results}.

At the peak time $t_2$ shown in Fig.~\ref{fig2}c,d, the EOVSA sources grow much brighter, reaching a brightness temperature of  $\sim3.3\times10^9$~K at the highest frequencies.  The weaker flank sources can no longer be seen, and an investigation of the time profiles of brightness of these sources shows that they intrinsically fade during the brightening of the central source, i.e. they do not merely become undetectable as a result of the $\sim$20:1 dynamic range that can be achieved in the EOVSA images.  By time $t_2$, the flux rope seen earlier in AIA has long-since left the field of view, but the strong energy release continues in the lower corona behind it.  The height dispersion of EOVSA source positions with frequency is similar to that at the earlier time, but the overall height of the sources evolves upward to remain well above the growing AIA bright loops.  The source shape at the highest frequency evolves toward a loop-like shape, but is asymmetric and slightly offset to the south.

The EOVSA source remains similar in shape to that in Fig.~\ref{fig2}c,d for 10s of minutes, slowly rising and growing weaker, but eventually it bifurcates and moves to the sides of the rising AIA bright loops, as shown at time $t_3$ in Fig.~\ref{fig2}e,f.  By this time, a bright ray has developed in AIA~193~\AA, which was studied in more detail by \citet{2018ApJ...854..122W} in conjunction with EUV spectral imaging data from the EUV Imaging Spectrometer \citep[EIS,][]{2007SoPh..243...19C} aboard Hinode \citep{2007SoPh..243....3K} and interpreted as a plasma sheet at a temperature of $\sim$20 MK. The EOVSA emission seems to avoid this location, and shows interesting frequency structure, with the southern source being stronger at high frequencies while the northern source dominates at low frequencies.  This likely reflects differences in both magnetic field structure and electron energy distributions on the two flanks.

\subsection{RHESSI Source Morphology}

Also shown in Figure~\ref{fig2} as open contours are the RHESSI HXR source locations, with red contours showing the lower-energy, thermal source and blue contours showing the higher-energy, nonthermal source. At each of the three times, the HXR thermal source is located within the bright AIA 193~\AA\ loops. The response function of the AIA  193~\AA\ band has a peak at $\sim$18 MK dominated by the highly ionized iron line Fe~XXIV  \citep{2010A&A...521A..21O}. The close spatial association between the HXR thermal source and the bright (or high emission measure) AIA 193~\AA\ loops suggest the presence of dense, superhot ($\sim$20~MK) plasma there, which was confirmed by the analysis of EIS data by \citet{2018ApJ...854..122W}.  At the initial time $t_1$, Fig.~\ref{fig2}a, a compact nonthermal footpoint HXR source is seen at the limb, which coincides with a bright kernel of white-light emission seen in continuum images from the Helioseismic and Magnetic Imager \citep[HMI,][]{2012SoPh..275..207S} at this time.  The conjugate footpoint HXR source is presumably hidden beyond the limb.  For similar events, see \citet{2015ApJ...802...19K}. A larger, more extended nonthermal HXR source is shown in Fig.~\ref{fig2}a at a position that agrees well with the MW emission above the AIA bright loops.  In order to image this weak nonthermal source, a two-step CLEAN procedure \citep{2011ApJ...742...82K} was used in which the brighter footpoint source was first imaged and subtracted from the HXR visibility data.  A second stage of CLEAN using the subtracted visibilities revealed the weaker source, which a series of imaging tests shows is quite robust.

At the peak time $t_2$ shown in Fig.~\ref{fig2}c,d, the RHESSI nonthermal 35-60 keV HXR source is found to be about 10\arcsec\ higher than the thermal 12-20 keV HXR source, but confined to the high-density region of the upper part of the AIA bright loops, at a projected height of $\sim$25~Mm.  Although RHESSI HXR data extend to still higher energies at this peak time, the high level of pulse pile-up means that imaging at higher energies requires further investigation, and therefore we do not show such higher-energy HXR images in Fig.\ref{fig2}c,d.

After time $t_2$, RHESSI entered the South Atlantic Anomally (SAA) and did not observe the Sun again until time $t_3$ shown in Fig.~\ref{fig2}e,f.  By this time, the AIA bright loops have grown to much greater heights ($\sim$45~Mm), and the nonthermal HXR emission extends above the densest part of the 193~\AA\ loops, encompassing the lower part of the bright ray and falling between the bifurcated MW sources.  The 10\% contour for the RHESSI nonthermal source is shown dashed, and indicates that the region of nonthermal HXR emission extends along the outside of the AIA bright loops, similar to the MW emission. 

\subsection{Temporal Development}
\begin{figure}    
   \centerline{\includegraphics[width=1.0\columnwidth,clip=]{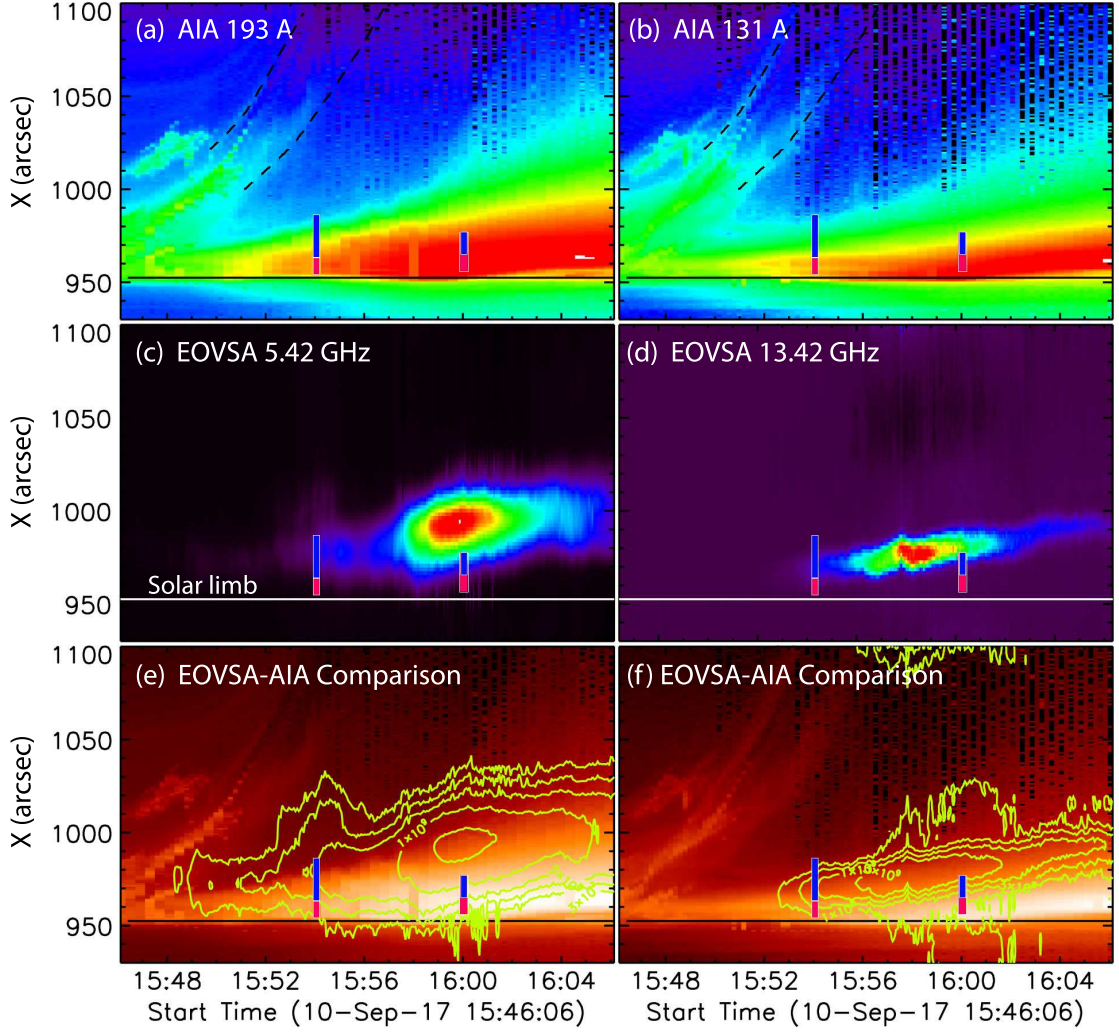}
              }
              \caption{Height-time stackplots of AIA, EOVSA, and RHESSI data from a horizontal cut at vertical position $y=-141\arcsec$ in Fig.~\ref{fig2}. The red vertical bars in each panel show the 50\% contour height range of the RHESSI thermal sources at times $t_1$ and $t_2$, from Fig.~\ref{fig2}a,c, while the blue bars show the corresponding height range of the RHESSI nonthermal sources. The time range covers 20 minutes, from 15:46-16:06 UT. (a) AIA 193 \AA\ intensity, log-scaled. The black dashed curves schematically show the leading and trailing edges of the flux rope. (b) AIA 131 \AA\ intensity, log-scaled. (c) EOVSA 5.42~GHz brightness temperature, linearly scaled. (d) EOVSA 13.42~GHz brightness temperature, linearly scaled. (e) The same AIA 193 \AA\ intensity as in (a) overlaid with the 5.42~GHz brightness temperature contours at 30, 100, 300, 1000, and 3000~MK. (f) The AIA 131 \AA\ intensity as in (b), overlaid with the 13.42 GHz brightness temperature contours at 100, 300, 1000, and 3000~MK.}
   \label{fig3}
\end{figure}

To visualize the temporal development of AIA, EOVSA, and RHESSI sources, we construct height-time stack plots in Figure~\ref{fig3}, along a cut taken parallel to the heliocentric $x$ axis at position $y = -141\arcsec$ bisecting the AIA loops.  The position of the cut is shown by the black dashed line in Fig.~\ref{fig2}d. The time resolution of the AIA data is 12 s, while the time resolution of the EOVSA images is 4 s (i.e. we made the EOVSA images at one-quarter of the available resolution of 1 s). The initial rise of the ejecta and flux rope during the first 10 minutes, which were studied by \citet{2018ApJ...853..178D} and  \citet{2018ApJ...855...74L}, manifests as an upward-moving feature apparent in the AIA 193 \AA\ and 131 \AA\ data, outlined with black-dashed curves in Fig.~\ref{fig3}a and \ref{fig3}b, respectively.  The slower rise of the newly formed EUV ``post-flare" loops occurs steadily throughout the period.  The corresponding EOVSA data for 5.42 and 13.42~GHz, shown in Fig.~\ref{fig3}c and \ref{fig3}d, respectively, also show the same steady rise in height at least after about 15:56~UT when the lower-frequency emission associated with the plasma sheet has faded.  Fig.~\ref{fig3}e,f repeat the AIA data from the upper panels, now overlaid with EOVSA contours from the middle panels to better demonstrate that the MW emission is located well above the EUV loops. In each panel, the height ranges of the RHESSI sources at the two times of Fig.~\ref{fig2}a,c are shown by the vertical bars, with red representing the thermal ($\sim$12~keV) source and blue the nonthermal ($\sim$35~keV) source.  The line labeled ``Solar limb" in Fig.~\ref{fig3}c marks the approximate height of the EUV limb.  The photosphere is $10\arcsec$ lower.  

\begin{figure}    
   \centerline{\includegraphics[width=1.0\columnwidth,clip=]{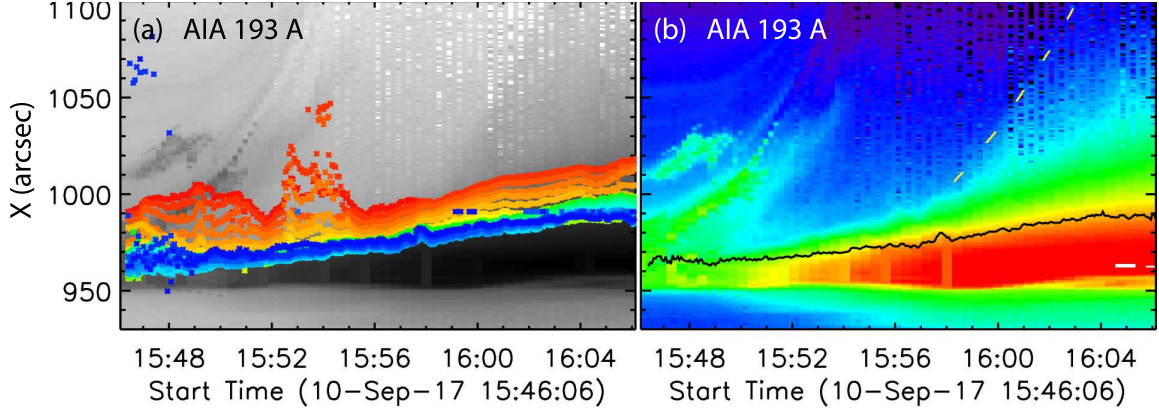}
              }
              \caption{(a) Height-time stackplot of AIA 193 \AA\, log-scaled in reverse gray scale (brightest emission is black), with EOVSA centroid source heights at all 30 frequencies, shown as symbols color-coded in frequency from red (spw 1 = 3.42~GHz) to blue (spw 30 = 17.92~GHz). (b) Repeat of log-scaled AIA 193 \AA\ height-time plot in rainbow colors, overplotted with the median centroid height (black curve) of EOVSA spw 26-30 (15.92-17.92~GHz).  The yellow dashed curve indicates the leading edge of the bright ray that grows rapidly along the plasma sheet.}
   \label{fig4}
\end{figure}

Figure~\ref{fig4}a better shows the frequency dependence of source height, where the symbols are color-coded in frequency from red (spectral window 1 = 3.42~GHz) to blue (spectral window 30 = 17.92~GHz).  The centroid source heights were determined by Gaussian fitting the one-dimensional profile of brightness temperature vs. height for each frequency and time.  The red symbols in Fig.~\ref{fig4}a show the tendency of the low-frequency EOVSA sources to follow the bright ejecta up to about 15:51~UT, after which the fading emission from the ejecta causes the source centroid to move back to the rising post-flare loops.  Then, from 15:52-15:55~UT, the low-frequency source extends upward again, along the plasma sheet below the rapidly rising flux rope.  This can also be seen in the contours of the 5.42~GHz source in Fig.~\ref{fig3}e.  After 15:55~UT, the MW emission at all frequencies settles into a slow increase in height with time, maintaining a dispersion of height with frequency.  The black curve in Figure~\ref{fig4}b shows the median height of the source in the frequency range 15.92-17.92~GHz, located near the top of the brightest EUV emission.  Remarkably, this source height closely tracks the AIA 193 \AA\ intensity contour.  The speed at the time of the most rapid rise, from 15:58:24-16:01:44~UT is $\sim$30~km~s$^{-1}$.  The more rapid rise in EUV indicated by the yellow dashed curve in Figure~\ref{fig4}b is due to previously-mentioned bright ray propagating outward at $\sim$288 km s$^{-1}$ along the plasma sheet, according to \citet{2018ApJ...854..122W}.

\section{Spectral Diagnostics} \label{sec:results}

\begin{figure}    
   \centerline{\includegraphics[width=1.0\columnwidth,clip=]{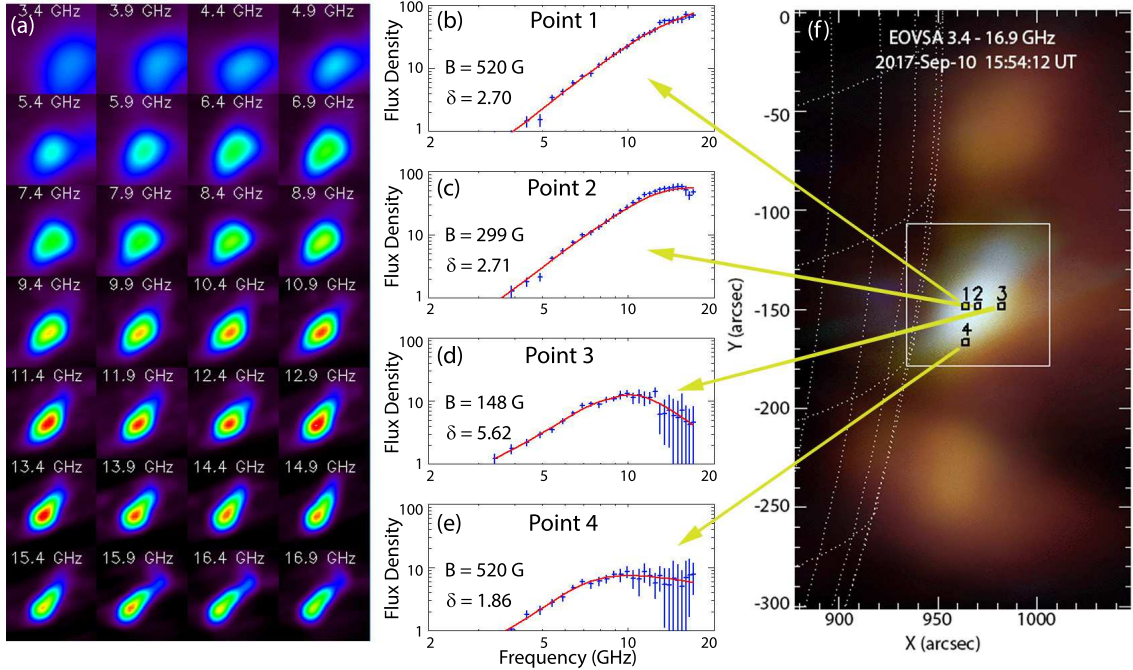}
              }
              \caption{The same analysis as for the simulations in \citet{2013SoPh..288..549G}, but with data from the time shown in Fig.~\ref{fig2}a,b. (a) Individual images at 28 frequencies, from the location of the white box in the overview image in panel f. (b-e) Measured flux-density spectra (points with $\pm1\sigma$ error bars) in single pixels of the images in panel a, corresponding to locations 1-4 marked in panel f, and corresponding multi-parameter fits (red lines). (f) A ``true-color'' representation of the EOVSA data cube, combining images at the 28 frequencies shown in panel a.
                      }
   \label{fig5}
\end{figure}

The EOVSA multi-frequency images form a four-dimensional data cube, two spatial, one spectral, and one temporal.  Prior to the completion of EOVSA, \citet{2013SoPh..288..549G} performed a quantitative simulation of a flaring loop and explored the diagnostic power of the technique of MW imaging spectroscopy.  Now, for the first time, we have actual EOVSA data that permit  the type of quantitative analysis simulated there. The approach is to obtain brightness temperature spectra over the frequency axis along different lines of sight in space, and then do a multi-parameter fit \citep{2009ApJ...698L.183F} assuming that the source is homogeneous along the line of sight.  A full analysis of the data in this manner is beyond the scope of this work, and will be published elsewhere. Here, we illustrate the procedure for four lines of sight at 15:54~UT, the time of the images shown in Fig.~\ref{fig2}a,b. Figure~\ref{fig5} shows the result, presented in the same format as in the \citet{2013SoPh..288..549G} paper, for comparison. We emphasize that these results are preliminary pending further refinement of the absolute flux calibration, which is now underway.

%
The same EOVSA data shown in Fig.~\ref{fig2}b are used to create a ``true-color" image in Fig.~\ref{fig5}f, where images at 28 frequencies are apportioned different red-green-blue weights according to their frequency.  EOVSA single-frequency images within the white box are shown in Fig.~\ref{fig5}a. The EOVSA source shape changes with frequency from a cusp-shaped source at mid-frequencies (e.g. the 7.9~GHz image in Fig.~\ref{fig5}a) evolving toward a more loop-like shape at higher frequencies (15.9~GHz image). These position and morphology changes with frequency correspond to position-dependent spectral shapes.  The spectra at the four locations indicated in Fig.~\ref{fig5}f are shown by the symbols in Fig.~\ref{fig5}b-e. The displayed spectra are scaled from the original brightness temperature units to solar flux units (sfu/pixel, where 1 sfu = $10^{-22}$ W m$^{-2}$ Hz$^{-1}$, and each map pixel has an area of $2\arcsec\times2\arcsec$). The $\pm1\sigma$ errors shown are based on residual fluctuations in a region of the maps away from any sources.  The relatively large error bars on the high-frequency spectral points for points 3 and 4 in Fig.~\ref{fig5}d,e reflect the fact that they come from low-brightness regions in the same map as the very bright source centered at point 1, demonstrating that the dynamic range of these images is about 20:1.

As shown in Fig.~\ref{fig5}f, the points 1-3 are located at different heights along the bisector of the EUV loops, while point 4 is at the same coronal height as point 1, but to the south edge of the source. The spectra in Fig.~\ref{fig5}b-d show that the peak frequency moves progressively to lower frequencies as the height increases, as would be expected for a decreasing magnetic field strength with height.  The spectrum at point 4 is similar to point 3, but seems to be flatter at high frequencies.  Using the homogeneous source multi-parameter fitting procedure for gyrosynchroton emission from an isotropic power-law distribution of electrons, described by \citet{2009ApJ...698L.183F}, we obtain the red curves, which are acceptable fits to the data points (reduced $\chi^2$ ranges from 0.2-0.5, which suggests that the error bars in Fig.~\ref{fig5}b-d may in fact over-estimate the variance in the data.). The two key parameters, the magnetic field strength $B$ and the power-law index of the electron energy distribution, $\delta$, are listed as text in each spectrum panel. As expected from the shift of the spectral peak with height in the corona, the derived magnetic field strength drops from 520 G at point 1 to 148 G at point 3.  It is interesting, however, that the magnetic field strength at point 4, which is at the same height as point 1,  is the same as at point 1, even though the spectral peak frequency is closer to that at point 3. The uncertainties in $B$ from the fitting procedure are relatively small, ranging from 10--15\%, but quantifying the systematic uncertainties requires modeling.  The power-law index is around $\delta = 2.7$ at points 1 and 2, and steepens to $\delta = 5.6$ at point 3, while the spectrum is extremely flat at point 4 with $\delta = 1.86$.  Again, the fitting uncertainties in $\delta$ are small, around 5\%, but systematic uncertainties remain to be quantified.  
%
%



\begin{figure}    
   \centerline{\includegraphics[width=0.6\columnwidth,angle=90]{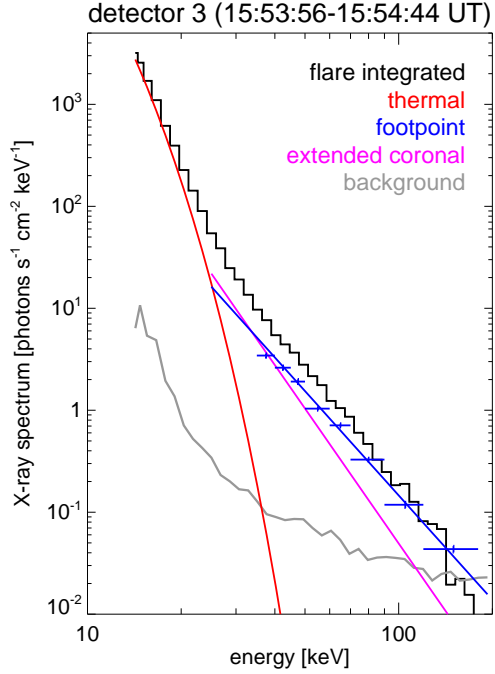}
              }
              \caption{RHESSI photon spectrum, where the black curve is the integrated spectrum, the red curve is the thermal source spectrum, and the blue points with a power-law fit are the energy distribution from the compact footpoint source. After subtraction, the implied spectrum of the extended coronal source at higher energies is shown as the purple line.  The power-law index of the compact source is $\gamma = 3.4$ while that of the extended source is $\gamma = 4.4$.
               }
   \label{fig6}
\end{figure}

It is useful to compare these derived parameters with spectral diagnostics from the RHESSI HXR data taken around 15:54~UT.  Figure~\ref{fig6} shows the results of such spectral analysis assuming a thermal-plus-single-power-law photon spectrum.  The flare-integrated HXR photon spectrum is shown by the black curve, which sums the contributions from the looptop thermal source, the compact footpoint source near the limb seen in Fig.~\ref{fig2}a, and the extended above-the-looptop nonthermal source.  For comparison with EOVSA, we are interested in this latter source, which is co-spatial with the EOVSA source region, but the HXR emission is too weak for accurate imaging spectroscopy.  However, the compact source is suitable for such imaging spectroscopy, which yields the blue crosses in Fig.~\ref{fig6}, and can be fit with a photon power-law index $\gamma = 3.4$ as shown by the blue line.  We then use the imaging spectroscopy result of the footpoint as a fixed input to the spatially-integrated spectral fitting, together with two free fit functions, a thermal component and a second power-law that represents the nonthermal coronal source. The fit to the nonthermal coronal source is shown in purple with a power-law index of $\gamma\approx4.4\pm0.1$. For the comparison with the radio derived spectral indexes we need the spectral index of the instantaneous distribution of nonthermal electrons, which comes from the thin-target model; thus, we get $\delta_{HXR} = \gamma-0.5 \approx 3.9$. 
The brightness center of the coronal HXR source is between points 2 and 3 in Figure~\ref{fig5}f; thus, having $\delta_{HXR}=3.9$ in between $\delta_{2}=2.7$ and $\delta_{3}=5.6$ derived from EOVSA looks reasonable.

\section{Discussion}
The combination of EUV, HXR, and MW imaging of the central source during the early impulsive phase (15:54~UT) matches expectations from the CSHKP model very well.  The bright EUV-emitting loops overlap the RHESSI thermal component, while the RHESSI nonthermal component comes from an extended region above them.  The EOVSA emission overlaps the RHESSI nonthermal source, but extends to greater heights at the lower MW frequencies where the magnetic field strength is lower. We have shown for the first time that it is possible to fit gyrosynchrotron spectra to spatially-resolved MW observations and derive a reliable set of physical parameters as a function of time and space. Spectral diagnostics of the accelerated electron spectrum derived from HXR and MW observations are broadly consistent.  A more detailed study is underway to exploit this technique to create dynamic parameters maps of the entire emitting region, as simulated by \citet{2013SoPh..288..549G}.

After the initial impulsive phase of the event, the closed-field region below the plasma sheet becomes the dominant region of MW emission, in agreement with the CSHKP model.  We expect that the bright and growing MW emission during this time is due to the continued creation of new loops combined with efficient electron trapping, which enables an increasing accumulation of high-energy electrons as the acceleration continues. Enhanced trapping due to an initially high mirror ratio \citep[e.g.][]{2004A&A...419.1159K} may explain the confinement of the MW sources to the loop tops.  For a static source, Coulomb collisions should lead to preferential loss of lower-energy particles, which would lead to progressive hardening and a positive delay of the MW peak with respect to frequency, which is opposite to the sense of delay shown in Fig.\ref{fig1}a,b.  However, in this event the source is not static, but growing upward as new loops are formed, which provides an opportunity for a more complex temporal evolution. The inferred energy of these MW-emitting electrons is extremely high.  The peak brightness temperature $T_b$ exceeds $3\times10^9$~K, which corresponds to a bulk electron energy for the emitting particles of $E > T_{\rm eff}/k \approx 270$~keV, where $T_{\rm eff} \approx T_b$ \citep[cf.][]{1985ARA&A..23..169D} is the effective electron temperature assuming optically-thick emission, and $k$ is the Boltzmann constant. The relatively flat energy distribution ($\delta = 2.7$) implies the presence of $>$1~MeV electrons in signficant numbers in the region, although quantitative estimates require additional analysis.

Energy release and acceleration of particles to high energies continues for hours after the event began, as revealed by EOVSA and by the continued level of $>$100~MeV emission detected by Fermi \citep{2018arXiv180307654O}. The EOVSA source becomes distinct from the HXR source during this time ($\sim$16:41~UT).  Both thermal and nonthermal RHESSI images match well the denser, EUV-emitting regions while the EOVSA source bifurcates and avoids the dense regions. Whether this is primarily due to suppression of the MW emission or a relative lack of higher energy electrons in the denser regions remains to be determined by the more-thorough spectral analysis now underway.

\section{Conclusion} \label{sec:conclusion}
We have used our coverage of the well-observed X8.2 limb flare, SOL2017-09-10, as an opportunity to present the first science results from a new, multi-frequency imaging array, the Expanded Owens Valley Solar Array. The results both agree with the standard CSHKP model for solar flares, and suggest the need for amending it, by revealing new details of the spatial distribution of high-energy electrons. MW observations at high, optically-thin frequencies provide source information in regions of high magnetic field strength, which are limited to relatively small, closed magnetic loops formed below the reconnection region.  The EOVSA images at lower MW frequencies early in the event reveal the prompt presence of high-energy electrons over a much larger region, including the plasma sheet extending between the lower, newly-formed loops and the rising flux rope, and the legs of a much larger loop well outside the traditionally-observed, reconnected loops.  Although isolated examples of such large source regions have been reported in the literature, as noted in section~\ref{sec:intro}, the ability of EOVSA to simultaneously image the whole MW spectrum, including both high- and low-frequency emission, has provided a panoramic view of the entire system of energetic electrons. Revealing the large spatial extent of the region of high-energy electrons is one of the key new insights provided by EOVSA, but equally important is its ability to provide quantitative diagnostics of plasma and particle parameters through MW imaging spectroscopy.  Further analysis of the dynamically evolving, spatially-resolved spectra is underway.

\acknowledgments
We thank the many talented engineers and technicians who worked to make EOVSA a reality, in particular our site manager Kjell Nelin.  This work was supported by NSF grants AST-1615807, AST-1735405, AGS-1654382, AGS-1723436, AGS-1817277 and NASA grants NNX14AK66G, 80NSSC18K0015, NNX17AB82G, NNX16AL67G, and 80NSSC18K0667 to New Jersey Institute of Technology. The RHESSI related part of this work is supported by NASA contract NAS 5-98033. SW acknowledges support from AFOSR LRIRs 14RV14COR and 17RVCOR416.

\bibliography{bibliography,2017sep10}

\begin{thebibliography}{}
\expandafter\ifx\csname natexlab\endcsname\relax\def\natexlab#1{#1}\fi
\providecommand{\url}[1]{\href{#1}{#1}}

\bibitem[{{Aschwanden} \& {Benz}(1997)}]{1997ApJ...480..825A}
{Aschwanden}, M.~J., \& {Benz}, A.~O. 1997, \apj, 480, 825

\bibitem[{{Atwood} {et~al.}(2009){Atwood}, {Abdo}, {Ackermann}, {Althouse},
  {Anderson}, {Axelsson}, {Baldini}, {Ballet}, {Band}, {Barbiellini}, \&
  et~al.}]{2009ApJ...697.1071A}
{Atwood}, W.~B., {Abdo}, A.~A., {Ackermann}, M., {et~al.} 2009, \apj, 697, 1071

\bibitem[{{Bastian} \& {Gary}(2005)}]{2005ASPC..345..142B}
{Bastian}, T.~S., \& {Gary}, D.~E. 2005, in Astronomical Society of the Pacific
  Conference Series, Vol. 345, From Clark Lake to the Long Wavelength Array:
  Bill Erickson's Radio Science, ed. N.~{Kassim}, M.~{Perez}, W.~{Junor}, \&
  P.~{Henning}, 142

\bibitem[{{Carmichael}(1964)}]{1964NASSP..50..451C}
{Carmichael}, H. 1964, NASA Special Publication, 50, 451

\bibitem[{{Culhane} {et~al.}(2007){Culhane}, {Harra}, {James}, {Al-Janabi},
  {Bradley}, {Chaudry}, {Rees}, {Tandy}, {Thomas}, {Whillock}, {Winter},
  {Doschek}, {Korendyke}, {Brown}, {Myers}, {Mariska}, {Seely}, {Lang}, {Kent},
  {Shaughnessy}, {Young}, {Simnett}, {Castelli}, {Mahmoud}, {Mapson-Menard},
  {Probyn}, {Thomas}, {Davila}, {Dere}, {Windt}, {Shea}, {Hagood}, {Moye},
  {Hara}, {Watanabe}, {Matsuzaki}, {Kosugi}, {Hansteen}, \&
  {Wikstol}}]{2007SoPh..243...19C}
{Culhane}, J.~L., {Harra}, L.~K., {James}, A.~M., {et~al.} 2007, \solphys, 243,
  19

\bibitem[{{Dennis}(1988)}]{1988SoPh..118...49D}
{Dennis}, B.~R. 1988, \solphys, 118, 49

\bibitem[{{Doschek} {et~al.}(2018){Doschek}, {Warren}, {Harra}, {Culhane},
  {Watanabe}, \& {Hara}}]{2018ApJ...853..178D}
{Doschek}, G.~A., {Warren}, H.~P., {Harra}, L.~K., {et~al.} 2018, \apj, 853,
  178

\bibitem[{{Dulk}(1985)}]{1985ARA&A..23..169D}
{Dulk}, G.~A. 1985, \araa, 23, 169

\bibitem[{{Fleishman} {et~al.}(2009){Fleishman}, {Nita}, \&
  {Gary}}]{2009ApJ...698L.183F}
{Fleishman}, G.~D., {Nita}, G.~M., \& {Gary}, D.~E. 2009, \apjl, 698, L183

\bibitem[{{Fleishman} {et~al.}(2015){Fleishman}, {Nita}, \&
  {Gary}}]{2015ApJ...802..122F}
---. 2015, \apj, 802, 122

\bibitem[{{Fleishman} {et~al.}(2017){Fleishman}, {Nita}, \&
  {Gary}}]{Fl_etal_2017}
---. 2017, \apj, 845, 135

\bibitem[{{Fletcher} \& {Martens}(1998)}]{1998ApJ...505..418F}
{Fletcher}, L., \& {Martens}, P.~C.~H. 1998, \apj, 505, 418

\bibitem[{{Gary} {et~al.}(2013){Gary}, {Fleishman}, \&
  {Nita}}]{2013SoPh..288..549G}
{Gary}, D.~E., {Fleishman}, G.~D., \& {Nita}, G.~M. 2013, \solphys, 288, 549

\bibitem[{{Gary} \& {Hurford}(1989)}]{Gary_Hurford_1989}
{Gary}, D.~E., \& {Hurford}, G.~J. 1989, \apj, 339, 1115

\bibitem[{{Gary} \& {Keller}(2004)}]{2004ASSL..314.....G}
{Gary}, D.~E., \& {Keller}, C.~U., eds. 2004, Astrophysics and Space Science
  Library, Vol. 314, {Solar and Space Weather Radiophysics - Current Status and
  Future Developments}

\bibitem[{{Gary} {et~al.}(2018){Gary}, {Chen}, {Grammer}, {Hickish}, {Hurford},
  {Lamb}, {MacMahon}, {McTiernan}, {Nelin}, {Nita}, \&
  {White}}]{2018SoPh...xx.xxxG}
{Gary}, D.~E., {Chen}, B., {Grammer}, W., {et~al.} 2018, \solphys, tbd

\bibitem[{{Guidice} \& {Castelli}(1975)}]{1975SoPh...44..155G}
{Guidice}, D.~A., \& {Castelli}, J.~P. 1975, \solphys, 44, 155

\bibitem[{{Hanaoka}(1997)}]{1997SoPh..173..319H}
{Hanaoka}, Y. 1997, \solphys, 173, 319

\bibitem[{{Hirayama}(1974)}]{1974SoPh...34..323H}
{Hirayama}, T. 1974, \solphys, 34, 323

\bibitem[{{Karlick{\'y}} \& {Kosugi}(2004)}]{2004A&A...419.1159K}
{Karlick{\'y}}, M., \& {Kosugi}, T. 2004, \aap, 419, 1159

\bibitem[{{Kopp} \& {Pneuman}(1976)}]{1976SoPh...50...85K}
{Kopp}, R.~A., \& {Pneuman}, G.~W. 1976, \solphys, 50, 85

\bibitem[{{Kosugi} {et~al.}(2007){Kosugi}, {Matsuzaki}, {Sakao}, {Shimizu},
  {Sone}, {Tachikawa}, {Hashimoto}, {Minesugi}, {Ohnishi}, {Yamada}, {Tsuneta},
  {Hara}, {Ichimoto}, {Suematsu}, {Shimojo}, {Watanabe}, {Shimada}, {Davis},
  {Hill}, {Owens}, {Title}, {Culhane}, {Harra}, {Doschek}, \&
  {Golub}}]{2007SoPh..243....3K}
{Kosugi}, T., {Matsuzaki}, K., {Sakao}, T., {et~al.} 2007, \solphys, 243, 3

\bibitem[{{Krucker} {et~al.}(2011){Krucker}, {Kontar}, {Christe}, {Glesener},
  \& {Lin}}]{2011ApJ...742...82K}
{Krucker}, S., {Kontar}, E.~P., {Christe}, S., {Glesener}, L., \& {Lin}, R.~P.
  2011, \apj, 742, 82

\bibitem[{{Krucker} \& {Lin}(2008)}]{2008ApJ...673.1181K}
{Krucker}, S., \& {Lin}, R.~P. 2008, \apj, 673, 1181

\bibitem[{{Krucker} {et~al.}(2015){Krucker}, {Saint-Hilaire}, {Hudson},
  {Haberreiter}, {Martinez-Oliveros}, {Fivian}, {Hurford}, {Kleint},
  {Battaglia}, {Kuhar}, \& {Arnold}}]{2015ApJ...802...19K}
{Krucker}, S., {Saint-Hilaire}, P., {Hudson}, H.~S., {et~al.} 2015, \apj, 802,
  19

\bibitem[{{Kucera} {et~al.}(1994){Kucera}, {Dulk}, {Gary}, \&
  {Bastian}}]{1994ApJ...433..875K}
{Kucera}, T.~A., {Dulk}, G.~A., {Gary}, D.~E., \& {Bastian}, T.~S. 1994, \apj,
  433, 875

\bibitem[{{Kuroda} {et~al.}(2018){Kuroda}, {Gary}, {Wang}, {Fleishman}, {Nita},
  \& {Jing}}]{2018ApJ...852...32K}
{Kuroda}, N., {Gary}, D.~E., {Wang}, H., {et~al.} 2018, \apj, 852, 32

\bibitem[{{Lee} {et~al.}(1994){Lee}, {Gary}, \& {Zirin}}]{1994SoPh..152..409L}
{Lee}, J.~W., {Gary}, D.~E., \& {Zirin}, H. 1994, \solphys, 152, 409

\bibitem[{{Lemen} {et~al.}(2012){Lemen}, {Title}, {Akin}, {Boerner}, {Chou},
  {Drake}, {Duncan}, {Edwards}, {Friedlaender}, {Heyman}, {Hurlburt}, {Katz},
  {Kushner}, {Levay}, {Lindgren}, {Mathur}, {McFeaters}, {Mitchell}, {Rehse},
  {Schrijver}, {Springer}, {Stern}, {Tarbell}, {Wuelser}, {Wolfson}, {Yanari},
  {Bookbinder}, {Cheimets}, {Caldwell}, {Deluca}, {Gates}, {Golub}, {Park},
  {Podgorski}, {Bush}, {Scherrer}, {Gummin}, {Smith}, {Auker}, {Jerram},
  {Pool}, {Soufli}, {Windt}, {Beardsley}, {Clapp}, {Lang}, \&
  {Waltham}}]{2012SoPh..275...17L}
{Lemen}, J.~R., {Title}, A.~M., {Akin}, D.~J., {et~al.} 2012, \solphys, 275, 17

\bibitem[{{Li} {et~al.}(2018){Li}, {Xue}, {Ding}, {Cheng}, {Su}, {Feng},
  {Hong}, {Li}, \& {Gan}}]{2018ApJ...853L..15L}
{Li}, Y., {Xue}, J.~C., {Ding}, M.~D., {et~al.} 2018, \apjl, 853, L15

\bibitem[{{Lin} {et~al.}(2002){Lin}, {Dennis}, {Hurford}, {Smith}, {Zehnder},
  {Harvey}, {Curtis}, {Pankow}, {Turin}, {Bester}, {Csillaghy}, {Lewis},
  {Madden}, {van Beek}, {Appleby}, {Raudorf}, {McTiernan}, {Ramaty}, {Schmahl},
  {Schwartz}, {Krucker}, {Abiad}, {Quinn}, {Berg}, {Hashii}, {Sterling},
  {Jackson}, {Pratt}, {Campbell}, {Malone}, {Landis}, {Barrington-Leigh},
  {Slassi-Sennou}, {Cork}, {Clark}, {Amato}, {Orwig}, {Boyle}, {Banks},
  {Shirey}, {Tolbert}, {Zarro}, {Snow}, {Thomsen}, {Henneck}, {McHedlishvili},
  {Ming}, {Fivian}, {Jordan}, {Wanner}, {Crubb}, {Preble}, {Matranga}, {Benz},
  {Hudson}, {Canfield}, {Holman}, {Crannell}, {Kosugi}, {Emslie}, {Vilmer},
  {Brown}, {Johns-Krull}, {Aschwanden}, {Metcalf}, \&
  {Conway}}]{2002SoPh..210....3L}
{Lin}, R.~P., {Dennis}, B.~R., {Hurford}, G.~J., {et~al.} 2002, \solphys, 210,
  3

\bibitem[{{Long} {et~al.}(2018){Long}, {Harra}, {Matthews}, {Warren}, {Lee},
  {Doschek}, {Hara}, \& {Jenkins}}]{2018ApJ...855...74L}
{Long}, D.~M., {Harra}, L.~K., {Matthews}, S.~A., {et~al.} 2018, \apj, 855, 74

\bibitem[{{Masuda} {et~al.}(1994){Masuda}, {Kosugi}, {Hara}, {Tsuneta}, \&
  {Ogawara}}]{1994Natur.371..495M}
{Masuda}, S., {Kosugi}, T., {Hara}, H., {Tsuneta}, S., \& {Ogawara}, Y. 1994,
  \nat, 371, 495

\bibitem[{{Meegan} {et~al.}(2009){Meegan}, {Lichti}, {Bhat}, {Bissaldi},
  {Briggs}, {Connaughton}, {Diehl}, {Fishman}, {Greiner}, {Hoover}, {van der
  Horst}, {von Kienlin}, {Kippen}, {Kouveliotou}, {McBreen}, {Paciesas},
  {Preece}, {Steinle}, {Wallace}, {Wilson}, \&
  {Wilson-Hodge}}]{2009ApJ...702..791M}
{Meegan}, C., {Lichti}, G., {Bhat}, P.~N., {et~al.} 2009, \apj, 702, 791

\bibitem[{{Melnikov} {et~al.}(2002){Melnikov}, {Shibasaki}, \&
  {Reznikova}}]{2002ApJ...580L.185M}
{Melnikov}, V.~F., {Shibasaki}, K., \& {Reznikova}, V.~E. 2002, \apjl, 580,
  L185

\bibitem[{{Nakajima} {et~al.}(1994){Nakajima}, {Nishio}, {Enome}, {Shibasaki},
  {Takano}, {Hanaoka}, {Torii}, {Sekiguchi}, {Bushimata}, {Kawashima},
  {Shinohara}, {Irimajiri}, {Koshiishi}, {Kosugi}, {Shiomi}, {Sawa}, \&
  {Kai}}]{1994IEEEP..82..705N}
{Nakajima}, H., {Nishio}, M., {Enome}, S., {et~al.} 1994, IEEE Proceedings, 82,
  705

\bibitem[{{Nishio} {et~al.}(1997){Nishio}, {Yaji}, {Kosugi}, {Nakajima}, \&
  {Sakurai}}]{1997ApJ...489..976N}
{Nishio}, M., {Yaji}, K., {Kosugi}, T., {Nakajima}, H., \& {Sakurai}, T. 1997,
  \apj, 489, 976

\bibitem[{{Nita} {et~al.}(2004){Nita}, {Gary}, \& {Lee}}]{2004ApJ...605..528N}
{Nita}, G.~M., {Gary}, D.~E., \& {Lee}, J. 2004, \apj, 605, 528

\bibitem[{{Nita} {et~al.}(2016){Nita}, {Hickish}, {MacMahon}, \&
  {Gary}}]{2016JAI.....541009N}
{Nita}, G.~M., {Hickish}, J., {MacMahon}, D., \& {Gary}, D.~E. 2016, Journal of
  Astronomical Instrumentation, 5, 1641009

\bibitem[{{O'Dwyer} {et~al.}(2010){O'Dwyer}, {Del Zanna}, {Mason}, {Weber}, \&
  {Tripathi}}]{2010A&A...521A..21O}
{O'Dwyer}, B., {Del Zanna}, G., {Mason}, H.~E., {Weber}, M.~A., \& {Tripathi},
  D. 2010, \aap, 521, A21

\bibitem[{{Omodei} {et~al.}(2018){Omodei}, {Pesce-Rollins}, {Longo},
  {Allafort}, \& {Krucker}}]{2018arXiv180307654O}
{Omodei}, N., {Pesce-Rollins}, M., {Longo}, F., {Allafort}, A., \& {Krucker},
  S. 2018, ArXiv e-prints, arXiv:1803.07654

\bibitem[{{Scherrer} {et~al.}(2012){Scherrer}, {Schou}, {Bush}, {Kosovichev},
  {Bogart}, {Hoeksema}, {Liu}, {Duvall}, {Zhao}, {Title}, {Schrijver},
  {Tarbell}, \& {Tomczyk}}]{2012SoPh..275..207S}
{Scherrer}, P.~H., {Schou}, J., {Bush}, R.~I., {et~al.} 2012, \solphys, 275,
  207

\bibitem[{{Schmahl} {et~al.}(1990){Schmahl}, {Schmelz}, {Saba}, {Strong}, \&
  {Kundu}}]{1990ApJ...358..654S}
{Schmahl}, E.~J., {Schmelz}, J.~T., {Saba}, J.~L.~R., {Strong}, K.~T., \&
  {Kundu}, M.~R. 1990, \apj, 358, 654

\bibitem[{{Sturrock}(1966)}]{1966Natur.211..695S}
{Sturrock}, P.~A. 1966, \nat, 211, 695

\bibitem[{{Wang} {et~al.}(1994){Wang}, {Gary}, {Lim}, \&
  {Schwartz}}]{1994ApJ...433..379W}
{Wang}, H., {Gary}, D.~E., {Lim}, J., \& {Schwartz}, R.~A. 1994, \apj, 433, 379

\bibitem[{{Wang} {et~al.}(2017){Wang}, {Liu}, {Ahn}, {Xu}, {Jing}, {Deng},
  {Huang}, {Liu}, {Kusano}, {Fleishman}, {Gary}, \&
  {Cao}}]{2017NatAs...1E..85W}
{Wang}, H., {Liu}, C., {Ahn}, K., {et~al.} 2017, Nature Astronomy, 1, 0085

\bibitem[{{Warren} {et~al.}(2018){Warren}, {Brooks}, {Ugarte-Urra}, {Reep},
  {Crump}, \& {Doschek}}]{2018ApJ...854..122W}
{Warren}, H.~P., {Brooks}, D.~H., {Ugarte-Urra}, I., {et~al.} 2018, \apj, 854,
  122

\bibitem[{{Yan} {et~al.}(2018){Yan}, {Yang}, {Xue}, {Mei}, {Kong}, {Wang}, \&
  {Li}}]{2018ApJ...853L..18Y}
{Yan}, X.~L., {Yang}, L.~H., {Xue}, Z.~K., {et~al.} 2018, \apjl, 853, L18

\end{thebibliography}




\end{document}